\documentclass[12pt]{article}
\begin{document}

\begin{center}
$\mathbf{A\;MODIFIED\;VARIATIONAL\;PRINCIPLE\;IN\;}$ $\mathbf{%
RELATIVISTIC\;HYDRODYNAMICS}$

{\bf I. Commutativity and Noncommutativity of the Variation Operator with
the Partial Derivative }

B.G.Dimitrov

{\it Joint Institute for Nuclear Research, Laboratory for Theoretical
Physics}

{\it Dubna 141 980, Russia}

E-mail: bogdan@thsun1.jinr.ru

{\bf Abstract}
\end{center}

A model is proposed, according to which the metric tensor field in the
standard gravitational Lagrangian is decomposed into a projection (generally
- with a non-zero covariant derivative) tensor field, orthogonal to an
arbitrary 4-vector field and a tensor part along the same vector field.
A theorem has been used, according to which the variation and the partial
derivative, when applied to a tensor field, commute with each other if and
only if the tensor field and its variation have zero covariant derivatives,
provided also the connection variation is zero. Since the projection field
obviously does not fulfill the above requirements of the ''commutation'' theorem,
the exact expression for the (non-zero) commutator of the variation and the
partial derivative, applied to the projection tensor field, can be found from
a set of the three defining equations.
The above method will be used to construct a modified variational approach in
relativistic hydrodynamics, based on variation of the vector field and the
projection field, the last one thus accounting for the influence of the reference
system of matter (characterized by the 4-vector) on the gravitational field.

\begin{center}
{\bf I. INTRODUCTION.}
\end{center}

The application of variational formalism in relativistic hydrodynamics has
always been a key problem for the adequate description of the combined
system of gravitational field and matter, understood in the sense of a
{\it perfect fluid}. In the notion of Dewitt [1], the perfect fluid is a
''stiff elastic'' medium. The most important physical quantities, allowing us
to view the perfect fluid (or matter) as a macro-body are the {\it %
mass-density }$\rho _o$, {\it the internal energy density }$w_o$ and
{\it the macrovelocity} $u$, the latter one making possible to
characterize the position of each particle of the ideal fluid. For all these
macrovariables the the corresponding conservation laws of {\it mass} and
{\it internal energy (or entropy}) can be defined. An important basic
point in the starting investigations [2,3] of relativistic hydrodynamics is
the following fact. Suppose the Einstein's equation is written as

\begin{equation}
R_{\mu \nu }-\frac 12g_{\mu \nu }R\equiv T_{\mu \nu }  \label{1},
\end{equation}
with the source term $T_{\mu \nu }$ in the right-hand side - {\it the
stress-energy tensor} of matter (in the case- the ideal fluid)

\begin{equation}
T_{\mu \nu }\equiv \rho _0u^\mu u^\nu -pg^{\mu \nu }  \label{2}.
\end{equation}
Then it can be proved that (1) and (2) are in fact the Euler's equations due
to the variation of $g_{\mu \nu }$ in the action functional

\begin{equation}
I=I_{matter}+I_{fluid}=\int \left[ R-2\rho _o(c^2+H_o+\frac 12\mu g_{\mu \nu
}u^\mu u^\nu )\sqrt{-g}d^4x\right]  \label{3},
\end{equation}
where $R$ is the scalar curvature of the gravitational field, $\mu $ is a
constant and
\begin{equation}
H_o=w_o-TS^o  \label{4}
\end{equation}
is the {\it Helmholz free energy, }$S_o$ is the {\it rest entropy} of
the fluid. However, this variational formalism suffers a defect - it is
based on the application of definite constraints, added in the standard
Gravitational Lagrangian in such a way that upon variation the {\it mass
conservation equation}

\begin{equation}
(\rho _ou^\nu )_{\mid \nu }\equiv 0  \label{5}
\end{equation}
($_{\mid \nu }$ means covariant differentiation)is fullfilled ($u^k$ is a
unit vector, such that $g_{\mu \nu }u^\mu u^\nu \equiv 1)$.In fact, the
addition of such constraints of a definite form is dictated by the necessity
to sideway the basic difference between general relativity and hydrodynamics
- while in general relativity all conservation laws are found on the base of
a {\it variational principle}{\bf , }in hydrodynamics these
conservation laws are of a ''continuim'' nature, in the sense that they are
a consequence of the conservation of mass and energy. Therefore, the
hydrodynamical equations, {\bf \ {\it are not derived on the base of a
variational principle.} }Note that in this formalism the equations of motion
of the fluid are nothing else but the equations of {\it energy-momentum
conservation}

\begin{equation}
T_{\mid \nu }^{\mu \nu }\equiv (R^{\mu \nu }-\frac 12g^{\mu \nu }R)_{\mid
\nu }  \label{6}.
\end{equation}
Also, it is evident that {\it the velocity field does not enter the
gravitational part of the Lagrangian.} Due to this reason the velocity field
is not subjected to a ''true'', physically reasonable variational principle.

In this paper another point of view shall be assumed - {\it if }$u^\alpha
(x)${\it \ is the attached to each space-time point of the fluid (medium)
velocity four-vector, then all physical vector and tensor variables in the
theory (for example - the metric tensor }$g_{\mu \nu }${\it ) should by
projected either along this velocity field,or along the perpendicular
direction}{\bf .} For example, if $e=u^ku_k$ is the velocity length
vector, then the tensor

\begin{equation}
p_{\mu \nu }\equiv g_{\mu \nu }-\frac 1eu_\mu u_\nu  \label{7}
\end{equation}
is called {\it a projection tensor}. Provided the vector field indices
are raised and lowered with the metric tensor, it can easily be checked that

\begin{equation}
u^\nu p_{\mu \nu }\equiv 0  \label{8}.
\end{equation}
{\it Therefore, at each spacetime point and at each instant of time the
vector field }$u${\it \ is orthogonal to the projection tensor. }Three
other facts are very important:

1. $u$ is an {\it arbitrary, not normalized, not time-like or space like
vector field.} In this sense, the transformation (7) can be viewed as a more
general one in comparison with the ADM approach [4], where $u$ was assumed
to be a timelike vector and the projection tensor $p_{\mu \nu }$-spacelike,
and also in comparison with the recently used in [5, 6] decomposition

\begin{equation}
\overline{g_{ij}}=g_{ij}+2Hl_il_j  \label{9},
\end{equation}
where $g_{ij},$ named a seed metric, was assumed to be a metric field (taken
to be conformally flat), and $l_i-$ a null vector field ($H=const.)$.

2. If $g_{\mu \nu}$is a metric tensor and $g^{\mu \nu }$ is its inverse,
then $p_{\mu \nu }$ {\it is not a metric tensor}, since from the defining
equation (7) it easily follows

\begin{equation}
p_{\mu \nu }p^{\nu \alpha }\equiv g_\mu ^\alpha -\frac 1eu^\alpha u_\mu
\label{10},
\end{equation}
where

\begin{equation}
g_\mu ^\alpha \equiv \delta _\mu ^\alpha \equiv \left\{ 1 { if }  \alpha
=\mu { and } 0 { if }\alpha \neq \mu \right\}   \label{11}.
\end{equation}

3. If $g_{\mu \nu }$ has a zero-covariant derivative with respect to the given
connection $\Gamma _{\mu \nu }^\alpha $, i.e.$\nabla _\alpha g_{\mu \nu
}\equiv 0$ ($\nabla _\alpha -$covariant derivative), then $p_{\mu \nu }$
does not have this property, because

\begin{equation}
\nabla _\alpha p_{\mu \nu }=-\nabla _\alpha (\frac 1eu_\mu u_\nu )
\label{12}.
\end{equation}
Evidently, this covariant derivative will be zero if $\nabla _\alpha u_\mu
=0 $, but the last would mean that a special kind of transport of the vector
field $u$ has been assumed. In fact, it turns out that the non-zero covariant
derivative of the tensor field $p_{\mu \nu }$ has profound consequences for the
variational formalism (called a {\it projection variational formalism}),
which will be based on variations of the projection field $p_{\mu \nu }$ and
the vector field $u$ and their (partial ) derivatives.This will be
further investigated.

The present paper is organized as follows: In Section II a proof will be
given (see [7] for the original version, also [8]) that the (form) variation
commutes with the partial derivative in the case of zero-covariant derivative
of the tensor field, provided also that the (form) variation of this field has also
zero-covariant dereivative and the (form) variation of the connection is zero. In
Section III a simple exaple will be given that in perturbative gravity
theory the (form)variation and the partial differentiation {\it do not
commute}. And as far as the consideration of the projection field as a
dynamical field variable is concerned, it is clear that {\it the
projection variational formalism strictly does not allow the implementation
of such a ''commutativity'' assumption due to the last property 3 and also
to the fact that the projection connection variation is different from zero
(even in case the initial connection variation is zero)}{\bf . }Due to
this reason, in Section III{\bf \ }the exact commutation relation between
the variation and the partial derivative, applied to the projection tensor,
will be found.

\begin{center}
{\bf II. WHEN\ DO\ THE\ VARIATION\ AND\ THE\ PARTIAL\ DERIVATIVE\ COMMUTE?%
}
\end{center}

In order to have an adequate understanding about the projection variational
formalism, one should first try to understand whether the commutation
property between the variation and the partial derivative, applied in the
theory of gravity for the derivation of the boundary terms for example, is
such a common property, which is suppposed to be applied without any
restrictions and assumptions.

The proposition below, originally proved in [7 ],will show that in fact the
commutation property places some restrictions and therefore it {\it %
cannot }always be applied. In fact, as noted in [8], in differentiable
manifolds {\it without affine connections} the ''commutativity'' property
is always fulfilled since the functional (form) variation is independent
from the change of coordinate ''maps''. But in differentiable manifolds with
affine connections, there is a relation between the functional (form)
variation and the covariant differentiation. This point of reasoning will be
explained in more details in the Discussion part of the present paper.

{\bf PROPOSITION 1 [9].} Let us denote by $P_{\mu \nu \alpha }$ the
following expression:

\begin{equation}
P_{\mu \nu \alpha }\equiv \overline{\delta }\partial _\alpha g_{\mu \nu
}-\partial _\alpha \overline{\delta }g_{\mu \nu }\equiv \left[ \overline{%
\delta },\partial _\alpha \right] g_{\mu \nu }  \label{13}
\end{equation}
and the variation is understood as a {\it form }variation $\overline{%
\delta }$, namely

\begin{equation}
\overline{\delta p}_{ij}\equiv p_{ij}^{^{\prime }}(x)-p_{ij}(x)  \label{14},
\end{equation}
thus representing the difference between the functional values, taken at one
and the same point.

Let us also assume that:

1. The covariant derivative of the background metric tensor in respect to
the background connection is zero, i.e. g$_{\mu \nu \mid \alpha }=0.$

2. The covariant derivative of the background metric tensor variation is
zero, i. e. $(\overline{\delta }g_{\mu \nu })_{\mid \alpha }\equiv 0$ and
therefore after the variation the metric remains again within the class of
Riemannian metrics with zero-covariant derivative.

Then $P_{ijk}\equiv 0$ if and only if $\overline{\delta }\Gamma _{\mu \nu
}^\alpha \equiv 0.$ In other words, {\it the variation commutes with the
partial derivative if and only if the variation of the background connection
is zero.}

{\bf Proof}: The proof is based essentially on the fact that the
variation $\delta g_{\mu \nu }$ is again a second - rank tensor and
therefore the standard formulae for its covariant derivative is valid:

\begin{equation}
(\overline{\delta }g_{\mu \nu })_{\mid \alpha }=\partial _\alpha \overline{%
\delta }g_{\mu \nu }+\Gamma _{\mu \alpha }^\sigma \overline{\delta }%
g_{\sigma \nu }+\Gamma _{\nu \alpha }^\sigma \overline{\delta }g_{\sigma \mu
}  \label{15}.
\end{equation}
Using the usual formulae for the covariant derivative $g_{\mu \nu \mid
\alpha }$and afterwards taking its variation, one can obtain:

\begin{equation}
\overline{\delta }(g_{\mu \nu \mid \alpha })=\overline{\delta }\partial
_\alpha g_{\mu \nu }+\overline{\delta }\Gamma _{\mu \alpha }^\sigma
g_{\sigma \nu }+\overline{\delta }\Gamma _{\nu \alpha }^\sigma g_{\sigma \mu
}+\Gamma _{\mu \alpha }^\sigma \overline{\delta }g_{\sigma \nu }+\Gamma
_{\nu \alpha }^\sigma \overline{\delta }g_{\sigma \mu }  \label{16}.
\end{equation}

From (15) and (16) it can be obtained:

\begin{equation}
P_{\mu \nu \alpha }=\overline{\delta }(g_{\mu \nu \mid \alpha })-(\overline{%
\delta }g_{\mu \nu })_{\mid \alpha }-\overline{\delta }\Gamma _{\mu \alpha
}^\sigma g_{\sigma \nu }-\overline{\delta }\Gamma _{\nu \alpha }^\sigma
g_{\sigma \mu }=  \label{17}
\end{equation}
\begin{equation}
=-\overline{\delta }\Gamma _{\mu \alpha }^\sigma g_{\sigma \nu }-\overline{%
\delta }\Gamma _{\nu \alpha }^\sigma g_{\sigma \mu }  \label{18},
\end{equation}

in accordance with the first and the second assumptions of this theorem.
From (17) it is clear that if $\overline{\delta }\Gamma _{\mu \nu }^\sigma
\equiv 0$, then $P_{\mu \nu \alpha }\equiv 0.$ Let us now assume that $%
P_{\mu \nu \alpha }\equiv 0.$ Then by cyclic permutation of the indices one
can obtain the relation:

\begin{equation}
0\equiv P_{\mu \nu \alpha }+P_{\nu \alpha \mu }-P_{\alpha \mu \nu }\equiv -2%
\overline{\delta }\Gamma _{\mu \alpha }^\sigma g_{\sigma \nu }  \label{19}
\end{equation}

and therefore $\overline{\delta }\Gamma _{\mu \alpha }^\sigma \equiv 0.$
This precludes the proof of the proposition.

It is clear from the above proof that the commutaion property will no longer
be valid for Einstein-Cartan theories, where the metric and the connection
are being varied independently. It is understood also that the requirement
about the zero connection variation is a more stronger one than the
requirement about the metric tensor, which means
that even within the class of Riemannian space-times the commutativity
between the variation and the partial derivative will not always be
fulfilled.

{\bf PROPOSITION\ 2}. Let us represent the metric tensor $g_{im}$ as

\begin{equation}
g_{im}=p_{im}+\frac 1eu_iu_m  \label{20}
\end{equation}
and let us also assume that the (form) variation and the partial derivative
commute in respect to the {\it metric} tensor $g^{im}$, i.e.

\begin{equation}
\left[ \overline{\delta },\partial _j\right] g^{im}\equiv 0  \label{21}.
\end{equation}
Then this relation, combined with the orthogonality condition

\begin{equation}
u^ip_{ik}\equiv 0  \label{22}
\end{equation}
leads to raising and lowering of the indices of the projection tensor with
the metric tensor, provided the indices of the vector field are also raised
and lowered with the metric tensor. Since this property is generally
fulfilled, in spite of any assumptions about commutativity or
non-commutativity, the statement here is that in fact the above property in
respect to the projection tensor indeed can be proved to follow from the
commutativity condition (21). In this sense, it does not provide any new
information.

{\bf Proof}: If (20) is substituted into (21), it can be obtained

\begin{equation}
\left[ \overline{\delta },\partial _j\right] p^{im}+\left[ \overline{\delta }%
,\partial _j\right] (\frac 1eu^iu^m)\equiv 0  \label{23}.
\end{equation}
Next, applying the commutator $\left[ \overline{\delta },\partial _j\right] $
to (22), written as $\frac 1eu^iu^mp_{ik}\equiv 0$, it can be derived that

\begin{equation}
\frac 1eu^iu^m\left[ \overline{\delta },\partial _j\right]
p_{ik}+p_{ik}\left[ \overline{\delta },\partial _j\right] (\frac
1eu^iu^m)\equiv 0  \label{24}.
\end{equation}
If the second term in (23) is substituted into (24), the following equation
is obtained

\begin{equation}
\frac 1eu^iu^m\left[ \overline{\delta },\partial _j\right]
p_{ik}-p_{ik}\left[ \overline{\delta },\partial _j\right] p^{im}\equiv 0
\label{25}.
\end{equation}
Taking into account also that
\begin{equation}
\frac 1eu^iu^m\left[ \overline{\delta },\partial _j\right] p_{ik}\equiv
g^{im}\left[ \overline{\delta },\partial _j\right] p_{ik}-p^{im}\left[
\overline{\delta },\partial _j\right] p_{ik}\equiv \left[ \overline{\delta }%
,\partial _j\right] p_k^m-p^{im}\left[ \overline{\delta },\partial _j\right]
p_{ik}  \label{26}
\end{equation}
{\it and making the very important assumption that the indices of the
projective tensor are raised and lowered with the metric tensor }$g_{ik}$,
equation (25) can be rewritten as

\begin{equation}
\left[ \overline{\delta },\partial _j\right] p_k^m\equiv p^{im}\left[
\overline{\delta },\partial _j\right] p_{ik}+p_{ik}\left[ \overline{\delta }%
,\partial _j\right] p^{im}\equiv \left[ \overline{\delta },\partial
_j\right] (p^{im}p_{ik})  \label{27},
\end{equation}
and from here
\begin{equation}
p_k^m\equiv p^{mi}p_{ik}  \label{28}.
\end{equation}

What is more important to be realized as the main result of this proposition
is the following: As a result of the assumption that indices of the
projection tensor can be raised and lowered with the given metric, (28) is
obtained. {\it As a matter of fact, however, this equation follows from
the defining equations (20) and (22) and also the assumption that the
indices of the vector field are raised and lowered also with the metric
tensor.} {\it In other words, the raising and lowering of projection
tensor indices with the metric tensor is not a direct consequence of the
commutativity property of the variation and the partial derivative in
respect to the metric tensor.} And of course, the last statement will also
be true if the vector field and the projection field interchange their
places.

\begin{center}
{\bf III. NONCOMMUTATIVITY\ OF\ THE\ VARIATION WITH\ THE\ PARTIAL\
DERIVATIVE IN\ PERTURBATIVE\ GRAVITY\ - A\ SIMPLE\ EXAMPLE}
\end{center}

In perturbative gravity the metric tensor $g_{\mu \nu }$ is expanded into
power series of the Planck constant $\hbar :$

\begin{equation}
g_{\mu \nu }\equiv g_{\mu \nu }^{(0)}+\hbar g_{\mu \nu }^{(1)}+\hbar
^2g_{\mu \nu }^{(2)}+.....=g_{\mu \nu }^{(0)}+\hbar \overline{\delta }g_{\mu
\nu }+\hbar ^2\overline{\delta }^2g_{\mu \nu }+...\equiv g_{\mu \nu
}^{(0)}+\hbar h_{\mu \nu }+\hbar ^2\overline{\delta }h_{\mu \nu }+...
\label{29},
\end{equation}
where the terms $\overline{\delta }g_{\mu \nu },$ $\overline{\delta }%
^2g_{\mu \nu }$ represent fluctuations of the gravitational field to first,
second and higher order. Note that while usually the background metric is
considered to have zero-covariant derivative, the fluctuations are thought
not to have this property. Therefore, after implementation of formulaes
(13) and (17), it can be derived

\begin{equation}
\left[ \overline{\delta },\partial _\alpha \right] g_{\mu \nu }\equiv
-h_{\mu \nu \mid \alpha }-\overline{\delta }\Gamma _{\alpha (\mu }^rg_{\nu
)r}  \label{30},
\end{equation}
where $\overline{\delta }(g_{\mu \nu \mid \alpha })\equiv 0$ and $\overline{%
\delta }g_{\mu \nu }\equiv h_{\mu \nu }$.

\begin{equation}
\left[ \overline{\delta },\partial _\alpha \right] h_{\mu \nu }\equiv
\overline{\delta }(h_{\mu \nu \mid \alpha })-(\overline{\delta }h_{\mu \nu
})_{\mid \alpha }-\overline{\delta }\Gamma _{\alpha (\mu }^rh_{\nu )r}\equiv
-(\overline{\delta }h_{\mu \nu })_{\mid \alpha }+\Gamma _{\alpha (\mu }^r%
\overline{\delta }h_{\nu )r}  \label{31}.
\end{equation}
The two formulaes clearly show that the variation and the partial derivative
{\it do not commute }neither when applied to the background metric, nor
to the fluctuating tensor field $h_{\mu \nu }$.

\begin{center}
{\bf IV. NON-COMMUTATIVITY\ BETWEEN\ THE\ VARIATION\ AND\ THE\ FIRST\
PARTIAL\ DERIVATIVE\ OF\ THE\ PROJECTION \ TENSOR - AN\ EXACT\ EXPRESSION\
FROM\ THE\ SET\ OF\ THREE\ DEFINING\ EQUATIONS.}
\end{center}

In the formal sense, a formulae analogous to (31) can also be written, but
for the projection field $p_{\mu \nu }.\,$In this section, however, the exact
expression for the commutator shall be derived on the basis of the three
equations (constraints), determining the projection metric in respect to the
vector field and the initially given metric field, namely:

1. The relation between the covariant and the contravariant components of
the projection tensor and the vector field, which is a consequence of the
existence of an inverse initial metric tensor $g^{\alpha \beta }:$

\begin{equation}  \label{32}
p_{mk}p^{ik}\equiv \delta _m^i-\frac 1eu^iu_m.
\end{equation}

2. The relation,expressing the orthogonality of the vector field $u$ in
respect to the projection tensor, which for convenience shall be written in
the form:

\begin{equation}  \label{33}
\frac 1eu^ku^ip_{km}\equiv 0.
\end{equation}

3. The relation, derived from the zero-covariant derivative of the
initial metric $g_{\alpha \beta }$ $($in respect to the initial
Christoffel connection $\Gamma _{\alpha \beta}^\gamma )$ , which
can be expressed as an nonlinear equation between the vector field $u$,
the projection tensor field $p_{\alpha \beta }$ and their
first partial derivatives

\begin{equation}
\partial _jg^{ki}\equiv -g^{s(k}\Gamma _{sj}^{i)}  \label{34},
\end{equation}

or after substituting $g^{ik}$ with $g^{ik}=p^{ik}+\frac 1eu^iu^k$:

\begin{equation}  \label{35}
\partial _jp^{ik}+\partial _j(\frac 1eu^iu^k)=-p^{s(k}\Gamma
_{sj}^{i)}-(\frac 1eu^su^{(k})\Gamma _{sj}^{i)},
\end{equation}

 where $\Gamma _{sj}^i$ is the connection of the inilially given metric.
In fact, this connection is a function of the projection tensor, the vector
field and their first partial derivatives and due to this (35) is a
nonlinear equation in respect to these variables. Note also the important
fact that (34) is a consequence of the fact that $g_{\alpha \beta }$ has a
well defined inverse metric $g^{\alpha \beta }.$

In order to find the exact formulae for the commutator $\left[ \delta
,\partial _j\right] p_{ik},$ the operators $\delta $ and $\partial _{j}$
shall be applied in a consequent and afterwards - in an inversed order to
all the above given relations (32), (33) and (35). Throughout the whole
section it shall be understood that when applied in a consequent order to a
given tensor field, first the action of the operator, standing left (i.e.
next) to the tensor, shall be assumed. For example, if we perform first
partial differentiation $\partial _j$ to equation (32) and afterwards - a
variation, we receive

\begin{equation}  \label{36}
p^{ik}(\delta \partial _jp_{km})+(\partial _jp_{mk})(\delta p^{ki})+(\delta
p_{mk})(\partial _jp^{ki})+p_{mk}(\delta \partial _jp^{ki})\equiv -\delta
\partial _j(\frac 1eu^iu_m).
\end{equation}
When applied in an inversed order, i.e. first the variation and then -the
partial differentiation, another equation is obtained

\begin{equation}
p^{ik}(\partial _j\delta p_{km})+(\partial _jp^{ik})(\delta
p_{km})+(\partial _jp_{mk})(\delta p^{ki})+p_{mk}(\partial _j\delta
p^{ki})\equiv -\partial _j\delta (\frac 1eu^iu_m)  \label{37}.
\end{equation}
From (36) and (37) the formulae for the commutator can be obtained

\begin{equation}
p^{ik}\left[ \delta ,\partial _j\right] p_{km}+p_{mk}\left[ \delta ,\partial
_j\right] p^{ki}\equiv -\left[ \delta ,\partial _j\right] (\frac 1eu^iu_m).
\label{38}
\end{equation}
In the same way, the commutator can be applied to (33) and it can be found

\begin{equation}
\frac 1eu^iu^k\left[ \delta ,\partial _j\right] p_{km}\equiv -p_{km}\left[
\delta ,\partial _j\right] (\frac 1eu^ku^i)  \label{39}.
\end{equation}
Having in mind also that indices of the vector field are lowered and lifted
with the initial metric and therefore $\frac 1eu^iu_m=\frac 1eu^ig_{mk}u^k$,
it can easily be derived that

\begin{equation}
\left[ \delta ,\partial _j\right] (\frac 1eu^iu_m)\equiv g_{mk}\left[ \delta
,\partial _j\right] (\frac 1eu^iu^k)+(\frac 1eu^iu^k)\left[ \delta ,\partial
_j\right] g_{mk}  \label{40}.
\end{equation}
If the right-hand side of (38) is substituted with the expression from the
last equation, the following equation is obtained:
\begin{equation}
p^{ik}\left[ \delta ,\partial _j\right] p_{km}+p_{mk}\left[ \delta ,\partial
_j\right] p^{ki}\equiv -g_{mk}\left[ \delta ,\partial _j\right] (\frac
1eu^iu^k)-(\frac 1eu^iu^k)\left[ \delta ,\partial _j\right] g_{mk}.
\label{41}
\end{equation}
Now use shall be made of the last relation (34) and the resulting equation
(35)$,$ following from the zero-covariant derivative of the initial metric.
Unlike the previously used equations (32) and (33), (35) includes in itself
terms with {\it partial derivatives} of the vector field and the
projection tensor. In other words, {\it the ''source'' of
noncommutativity is hidden in the fact that the partial differentiation does
not enter on an equal footing in all the three equations}. This means that
since partial differentiation is already present in (35), only the variation
can be performed

\begin{equation}
\delta \partial _j(\frac 1eu^ku^i)\equiv -\delta \partial _jp^{ik}-\delta
p^{s(k}\Gamma _{sj}^{i)}-p^{s(k}\delta \Gamma _{sj}^{i)}-\delta (\frac
1eu^su^{(k})\Gamma _{sj}^{i)}-\frac 1eu^su^{(k}\delta \Gamma _{sj}^{i)}
\label{42},
\end{equation}
and therefore, we have no longer a commutation relation for the two
operators in this particular equation.

Note that 1. The derivation of the last equation is crucial for proving the
noncommutativity property of the variation and the partial differentiation
in respect to the projection tensor. The previously derived equations (38)
and (39) {\bf cannot }by themselves prove this property. 2. For the moment%
{\it \ }we have not made use of any assumption, concerning {\it %
commutativity} of the partial differentiation with the variation in respect
to the {\it initially given metric tensor} $g_{\alpha \beta }$.
Therefore, the variation of the initial Christoffel connection is assumed to
be different from zero, and the same applies also to the last term in (40).

Now, the last expression (42) can be substituted into the first term on the
right-hand side of (41) to give

\[
g^{ik}\left[ \delta ,\partial _j\right] p_{mk}+p_{mk}\left[ \delta ,\partial
_j\right] p^{ik}\equiv g_{mk}\partial _j\delta (\frac 1eu^iu^k)+g_{mk}\delta
\partial _jp^{ik}+
\]

\[
+g_{mk}\delta p^{s(k}\Gamma _{sj}^{i)}+g_{mk}p^{s(k}\delta \Gamma
_{sj}^{i)}+g_{mk}\delta (\frac 1eu^su^{(k})\Gamma _{sj}^{i)}+
\]

\begin{equation}
+g_{mk}(\frac 1eu^su^{(k})\delta \Gamma _{sj}^{i)}-(\frac 1eu^iu^k)\left[
\delta ,\partial _j\right] (\frac 1eu_mu_k)  \label{43}.
\end{equation}

Also, the orthogonality property (37) can be written for the contravariant
tensor $p^{ik}$ in the form $\frac 1eu_ku_mp^{ki}\equiv 0$, and the
analogous to (38) expression for the commutator is

\begin{equation}
\frac 1eu_ku_m\left[ \delta ,\partial _j\right] p^{ik}\equiv -p^{ik}\left[
\delta ,\partial _j\right] (\frac 1eu_ku_m)  \label{44}.
\end{equation}
Summing up the last two equations (43) and (44) and contracting the
resulting equation with $g_{ir}$, it can be received, after some
transformations:

\begin{eqnarray}
\left[ \delta ,\partial _j\right] p_{mr} &\equiv &g_{ir}g_{mk}\partial
_j\left[ \delta p^{ik}+\delta (\frac 1eu^iu^k)\right] +\partial _j\delta
p_{mr}+  \nonumber  \label{45} \\
&&\ \ +g_{k(r}\left[ p^{si}g_{m)i}+\frac 1eu_{m)}u^s\right] \delta \Gamma
_{sj}^k+\frac 1eg_{l(r}u_{m)}\Gamma _{kj}^l\delta u^k+  \nonumber \\
&&\ \ +\Gamma _{sj}^lg_{l(r}\left[ \frac 1eu_ku_{m)}\delta _n^s+\frac
1eu^su_ng_{m)k}\right] \delta p^{nk}-  \nonumber \\
&&\ \ -g_{l(r}g^{np}p_{pm)}\Gamma _{sj}^lg^{sk}\delta p_{nk}-\frac
1{e^2}\Gamma _{nj}^lg_{l(r}u_{m)}u^nu^k\delta u_k{.}  \label{45}.
\end{eqnarray}
Note that this expression does not contain the commutator $\left[ \delta
,\partial _j\right] p^{ik}$, since it has cancelled after the summation of
(43) and (44). Also, it has been assumed that the variation and the partial
derivative commute in respect to the covariant vector field

\begin{equation}
\left[ \delta ,\partial _j\right] u_m=0  \label{46}.
\end{equation}
Note, however that in respect to the {\it contravariant vector field }the
following relation is fullfilled

\begin{equation}
\left[ \delta ,\partial _j\right] u^m=u_k\left[ \delta ,\partial _j\right]
g^{mk}+g^{mk}\left[ \delta ,\partial _j\right] u_k  \label{47}.
\end{equation}
{\it In other words, the variation and the partial derivative commute in
respect to the contravariant vector field if and only if they commute in
respect to the covariant vector field (equation (46) and also to the metric
tensor }$g_{ik}${\it , i.e. }$\left[ \delta ,\partial _j\right] g^{mk}=0$.
Relation (45) for the variation $\left[ \delta ,\partial _j\right] p_{mr}$
can be additionally simplified if all the variations of contravariant
quantities will be found and expressed through the variations of the
covariant ones. However, even after that the commutator (45) remains
non-zero.

\begin{center}
{\bf V. DISCUSSION\ AND\ CONCLUSIONS.}
\end{center}

In this paper a proposition has been used, which states under what
conditions the partial derivative and the variation commute with each other
- zero-covariant derivative of the tensor field and its form variation, and
also zero connection variation. The essence of the non-commutation (or commutation)
property can be understood if one tries to realize what is the effect of the
two differential operations- the variation and the partial differentiation,
when applied in a different order. Let us take first the partial
differentiation. According to the standard definition, it is merely the
difference between the functional values of the tensor field components at
two different spacetime points. However, the functional values are compared
after they ''resume'' the previous (i.e. initial) space-time point (but the
two functional values remain of course different). However, in a curved
space-time the process of determining the functional value differences is
possible only if the {\it connection variation }from (space-time) point
to point is given. Acting subsequently with the (form) variation operator,
it is understood that the variation acts also on the connection itself. To
put it into another way, this means that the form-variation is determined by
the connection variation because the variation is performed {\it after }%
partial differentiation - an operation, accounted by the functional tensor
components (form) variations, but depending in fact also on the space-time
properties. Now, if the form variation is first performed, by the nature of
its mathematical definition, {\it it doesn't account for the presence of
a curved space-time due to the simple reason that in no way the connection
can enter in the functional variation (unlike the previous case)}. In
both cases, when partial differentiation is performed in different
order, this differentiation is applied to tensor field components with
{\it different functional values, }which evolve in a different and
complicated manner from one space-time point to another. This explains why
the different consequitive{\it \ }applications of these operators leads
to different results.

Finally, since the ''commutation'' conditions are evidently not fulfilled
for the projection tensor field, in Section IV the exact expression (45) for
the commutator has been found on the base of the three defining equations.
It is important to mention, however, that in fact we have two expressions
for this commutator - the first is the already mentioned eq. (45) and the
other one is the derived in Section IV\ formulae (31), but for the
projection tensor

\begin{equation}
\left[ \overline{\delta },\partial _\alpha \right] p_{\mu \nu }\equiv
\overline{\delta }(p_{\mu \nu \mid \mid \alpha })-(\overline{\delta }p_{\mu
\nu })_{\mid \overline{\mid }\alpha }-\overline{\delta }\overline{\Gamma }%
_{\alpha (\mu }^rp_{\nu )r}  \label{48},
\end{equation}

where the double symbol $\mid \mid $ denotes covariant differentiation with
respect to the projection connection $\overline{\Gamma }_{\alpha \mu }^r$ ,
defined in the standard way

\begin{equation}
\overline{\Gamma }_{\alpha \mu }^r\equiv \frac 12p^{rs}(\partial _\alpha
p_{\mu s}+\partial _\mu p_{\alpha s}-\partial _rp_{\alpha \mu })
\label{49}.
\end{equation}
Note also that the projection connection is determined through a projection
tensor field, which does not have an inverse one and this case is different
from the standard Christoffell connection.

The last expression (48) is evidently related to performing a covariant
differentiation in respect to the projection connection and therefore - to
the kinds of transports in a space-time, determined by this connection.
However, if one has a classification of the kinds of transports in a
space-time with the initial connection, this doesn't mean that such a
classification is available also for a space time with the projection
connection. That is the reason in the present paper formulae (45) has been
used as the more preferred and suitable one and not (48). A problem for
further investigation remains open and without an answer yet - whether and
under conditions the two expressions can be equivalent, if this is at all
possible.

\begin{center}
{\bf Acknowledgements}
\end{center}

The author is grateful to Dr.S. Manoff from the Institute of Nuclear
Energetics and Nuclear Research of the Bulgarian Academy of Sciences for
useful discussions,critical remarks and advices, also for reading the
manuscript.

The author is grateful also to his colleagues P. Bojilov and D. Mladenov for
useful discussions and remarks, also to Prof. B. M. Barbashov from the
Laboratory for Theoretical Physics at the Joint Institute for Nuclear
Research (Dubna) for his interest towards the investigated problem.

This work has been supported partly by Grant No.F-642 of the National
Science Foundation of Bulgaria.

\begin{center}
{\bf REFERENCES}
\end{center}

[1] B. S. DeWitt, in: \emph{Gravitation: an introduction to current research}
(John Wiley \& Sons 1962)

[2] A. H. Taub, \emph{Phys. Rev}. {\bf 94}, 1468 (1969)

[3] A. H. Taub, \emph{Commun. Math. Phys}. $\mathbf{15}$, 235 (1969)

[4] R. Arnowitt, S. Deser and C. Misner, in: \emph{Gravitation: an
introduction to current research} (John Wiley \& Sons 1962)

[5] N. Dadhich, \emph{Isothermall spherical perfect fluid model: uniqueness
and conformal mapping}, Report gr-qc/9605002

[6] N. Dadhich, \emph{Conformal Mapping and Isothermal Perfect Fluid Model},
Report gr-qc/9605003

[7] S. Manoff, in \emph{Topics in Complex Analysis, Differential Geometry
and Mathematical Physics }, eds. S.Dimiev and K.\thinspace Sekigawa (World
Scientific, Singapore, 1997) ; \emph{Intern. Journ. Mod. Phys}. {\bf A13}%
, 1941 (1998)

[8] S. Manoff, A. Kolarov, B. G. Dimitrov, \emph{Commun. Joint Inst. Nucl.
Res.} {\bf E5-98-184}, Dubna, 1998 ; S. Manoff, \emph{Acta\thinspace
Applicandae Mathematicae} {\bf 55}, 51 (1999)

[9] S. Manoff, private communication

\end{document}